# EMOTION DETECTION FROM TEXT

Shiv Naresh Shivhare[1] and Prof. Saritha Khethawat[2]

[1]Department of CSE and IT, Maulana Azad National Institute of Technology, Bhopal, Madhya Pradesh, India
Shiv827@gmail.com
[2]Department of CSE and IT, Maulana Azad National Institute of Technology, Bhopal, Madhya Pradesh, India
sarithakishan@gmail.com

*ABSTRACT*

*Emotion can be expressed in many ways that can be seen such as facial expression and gestures, speech and by written text. Emotion Detection in text documents is essentially a content – based classification problem involving concepts from the domains of Natural Language Processing as well as Machine Learning. In this paper emotion recognition based on textual data and the techniques used in emotion detection are discussed.*

## 1. INTRODUCTION

Detecting emotional state of a person by analyzing a text document written by him/her appear challenging but also essential many times due to the fact that most of the times textual expressions are not only direct using emotion words but also result from the interpretation of the meaning of concepts and interaction of concepts which are described in the text document. Recognizing the emotion of the text plays a key role in the human-computer interaction [1]. Emotions may be expressed by a person's speech, face expression and written text known as speech, facial and text based emotion respectively. Sufficient amount of work has been done regarding to speech and facial emotion recognition but text based emotion recognition system still needs attraction of researchers [14]. In computational linguistics, the detection of human emotions in text is becoming increasingly important from an applicative point of view.

Emotion is expressed as joy, sadness, anger, surprise, hate, fear and so on. Since there is not any standard emotion word hierarchy, focus is on the related research about emotion in cognitive psychology domain. In 2001, W. Gerrod Parrot[2], wrote a book named "Emotions In Social Psychology", in which he explained the emotion system and formally classified the human emotions through an emotion hierarchy in six classes at primary level which are Love, Joy, Anger, Sadness, Fear and Surprise. Certain other words also fall in secondary and tertiary levels. Directions to improve the capabilities of current methods of text-based emotion detection are proposed in this paper.

## 2. RELATED WORK

The concept of affective computing in 1997 by Since Picard [3] proposed that the role of emotions in human computer interaction. This domain attracted many researchers from computer science, biotechnology, psychology, and cognitive science and so on. Following the trend, the research in the field of emotion detection from textual data emerged to determine human emotions from another point of view. Problem of emotion recognition from text can be formulated as follows: Let E be the set of all emotions, A be the set of all authors, and let T be the set of all possible representations of emotion-expressing texts. Let r be a function to reflect emotion e of author a from text t, i.e., r: A x T →E, then the function r would be the answer to the problem [4].

The main problem of emotion recognition systems lies in fact that, although the definitions of E and T may be straightforward, the definitions of individual element, even subsets in both sets of E and T would be rather confusing. On one side, for the set T, new elements may add in as the languages are constantly emerging. Whereas on the other side, currently there are no standard classifications of "all human emotions" due to the complex nature of human minds, and any emotion classifications can only be seen as "labels" annotated afterwards for different purposes. Methods used for text based emotion recognition system [4], [5] are:

## 2.1. Keyword Spotting Technique

The keyword pattern matching problem can be described as the problem of finding occurrences of keywords from a given set as substrings in a given string [4]. This problem has been studied in the past and algorithms have been suggested for solving it. In the context of emotion detection this method is based on certain predefined keywords. These words are classified into categories such as disgusted, sad, happy, angry, fearful, surprised etc. Process of Keyword spotting method is shown in the figure 1.

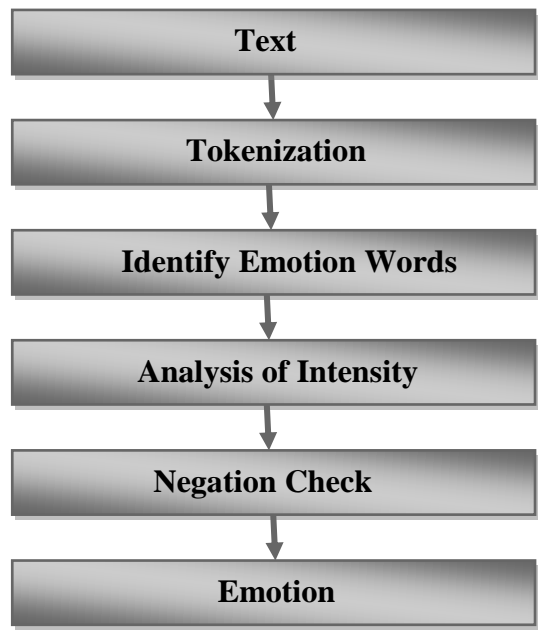

Figure 1. Keyword Spotting Technique

Keyword spotting technique for emotion recognition consists of five steps shown in fig.1 where a text document is taken as input and output is generated as an emotion class. At the very first step text data is converted into tokens, from these tokens emotion words are identified and detected. Initially this technique will take some text as input and in next step we perform tokenization to the input text. Words related to emotions will be identified in the next step afterwards analysis of the intensity of emotion words will be performed. Sentence is checked

whether negation is involved in it or not then finally an emotion class will be found as the required output.

## 2.2. Lexical Affinity Method

Detecting emotions based on related keywords is an easy to use and straightforward method. *Lexical Affinity approach* is an extension of keyword spotting technique; it assigns a probabilistic 'affinity' for a particular emotion to arbitrary words apart from picking up emotional keywords. These probabilities are often part of linguistic corpora, but have disadvantages; firstly the assigned probabilities are biased toward corpus-specific genre of texts, secondly it misses out emotional content that resides deeper than the word-level on which this technique operates. For example the word 'accident', having been assigned a high probability of indicating a negative emotion, would not contribute correctly to the emotional assessment of phrases like 'I avoided an accident' or 'I met my girlfriend by accident'.

## 2.3. Learning-based Methods

Learning-based methods are being used to formulate the problem differently. Originally the problem was to determine emotions from input texts but now the problem is to classify the input texts into different emotions. Unlike keyword-based detection methods, learning-based methods try to detect emotions based on a previously trained classifier, which apply various theories of machine learning such as support vector machines [8] and conditional random fields [9], to determine which emotion category should the input text belongs.

## 2.4. Hybrid Methods

Since keyword-based methods with thesaurus and naïve learning-based methods could not acquire satisfactory results, some systems use hybrid approach by combining both keyword spotting technique and learning based method, which help to improve accuracy. The most significant hybrid system so far is the work of Wu, Chuang and Lin [11], that utilizes a rule-based approach to extract semantics related to specific emotions and Chinese lexicon ontology to extract attributes. These semantics and attributes are associated with emotions in the form of emotion association rules. As a result, these emotion association rules, replacing original emotion keywords, serve as the training features of their learning module based on separable mixture models. This method outperforms previous approaches, but categories of emotions are still limited.

## 2.5. Limitations

From above discussion there are few limitations [7]:

### 2.5.1. Ambiguity in Keyword Definitions

Using emotion keywords is a straightforward way to detect associated emotions, the meanings of keywords could be multiple and vague, as most words could change their meanings according to different usages and contexts. Moreover, even the minimum set of emotion labels (without all their synonyms) could have different emotions in some extreme cases such as ironic or cynical sentences.

### 2.5.2. Incapability of Recognizing Sentences without Keywords

Keyword-based approach is totally based on the set of emotion keywords. Therefore, sentences without any keyword would imply that they do not contain any emotion at all, which is obviously wrong. For example, "I passed my qualify exam today" and "Hooray! I passed my qualify exam today" should imply the same emotion (joy), but the former without "hooray" could remain undetected if "hooray" is the only keyword to detect this emotion.

### 2.5.3. Lack of Linguistic Information

Syntax structures and semantics also have influences on expressed emotions. For example, "I laughed at him" and "He laughed at me" would suggest different emotions from the first person's perspective. As a result, ignoring linguistic information also poses a problem to keyword-based methods.

### 2.5.4. Difficulties in Determining Emotion Indicators [10]

Learning-based methods can automatically determine the probabilities between features and emotions but the methods still need keywords, but in the form of features. The most intuitive features may be emoticons which can be seen as author's emotion annotations in the texts. The cascading problems would be the same as those in keyword-based methods.

## 3. PROPOSED ARCHITECTURE

Methods described in section II are modified and integrated to extend their capabilities and to improve the performance for which a simple and easy to understand model is designed shown in fig.2.

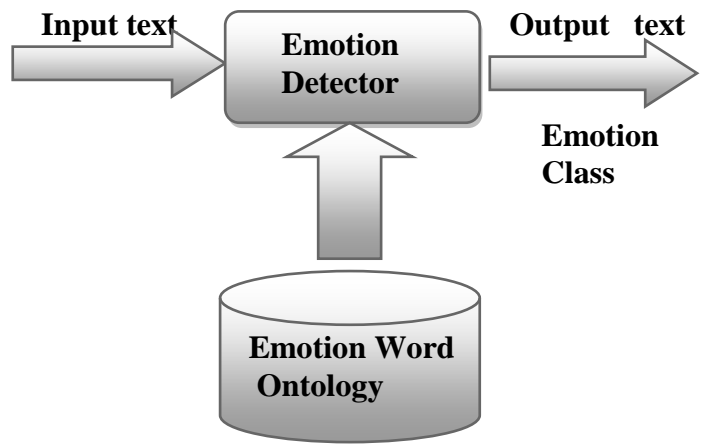

Figure 2. Proposed Architecture

The Framework is divided into two main components: Emotion Ontology, Emotion Detector.

### 3.1. Emotion Ontology

Ontology is an explicit specification of conceptualization. Ontologies have definitional aspects like high level schemas and aspects like entities and attributes [16]; interrelationship is between entities, domain vocabulary. Ontologies provide an understanding of particular domain. Ontologies allow the domain to be communicated between persons, institutions, and application systems. Emotion word hierarchy is converted into ontology. This emotion word hierarchy is developed by W.G. parrot. Protégé [13], an ontology development tool is used to develop emotion ontology. Proposed ontology has class and subclass relationship format. Emotion classes at the primary level in emotion hierarchy are at the top of emotion ontology and emotion classes at the tertiary level are at the bottom of ontology. High weight age is assigned to the upper level emotion classes and low to the lower level emotion classes.

## 3.2. Emotion Detector Algorithm

Emotion of the textual data can be recognized with the help of this emotion detection algorithm. The algorithm calculates weight for particular emotion by adding weights assigned at each level of hierarchy and also calculates same for its counter emotion, then compares the both scores and greater one is taken as the detected emotion.

### 3.2.1. Parameters Used

Algorithm is to calculate weight age to be assigned to different emotion words so that they can be sorted according to it. Certain parameters are required for this purpose. The first step is calculation of parameters. This task is achieved with the help of Jena library which allows traversal and parsing of ontology.
Different parameters are calculated as follows:

#### 3.2.1.1. Parent-Child relationship

If a text document belongs to a child; it also indirectly refers to the parent of it. Hence if a certain value is added to the child's score, parent score also need to be modified. This is achieved by traversing the ontology model in a breadth first manner using Jena API. When any node is encountered all of its children are retrieved. Then same method is applied to every child.

#### 3.2.1.2. Depth in Ontology

This is required as it gives an idea about how specific is the term in relation to its corresponding ontology structure. The more specific it is the more weight age should be given to it. This value is calculated simultaneously while traversing the ontology tree.

#### 3.2.1.3. Frequency in Text document

This is also an important parameter as more is the frequency more will be the importance of that term. This value is calculated by parsing the text document and searching for occurrences of the words.

### 3.2.2. Algorithm

Following algorithm is proposed to calculate the score for each emotion word with the help of parameters from previous steps. This score will be directly proportional to the frequency of the term and inversely proportional to its depth in the ontology. Hence a formula devised for the $m^{th}$ terminology. For every primary level emotion class, a respective score will be calculated. Finally Emotion class having highest score will win the race and declared as Emotion state of the corresponding text document. Algorithm is as follows

```
for j ← 1 to No. of Nodes [Ontology]
    do  parent [j] ← parent of node j
        child [j] ← child of node j
    for m ← 1 to No. of Nodes [Ontology]
        do freq [m] ← frequency of occurrence of m^th
           depth [m] ← depth of m^th node in ontology
```

**Calculate (x):**

```
for m ← 1 to No. of Nodes [Ontology]
    score (x) ← freq [root] / depth [root]
    for m ← 1 to No. of parent nodes [Ontology]
```

```
        score (parent) = score (parent) + score (child)
        return score (parent)
    for m ← 1 to No. of parent nodes [ontology]
emotion class ← High score [parent]
    return emotion class
```

*Where* Nodes [Ontology] denotes classes in the ontology, Parent [j] denotes parent classes in the ontology, Child [j] denotes child classes in the ontology, Freq [m] denotes frequency of occurrence of $m^{th}$ class in text, Depth denotes depth of class into ontology, Score [parent] denotes score of parent in ontology.

By proposed algorithm we can find out the score of primary emotion classes. Emotion class with highest score will be decided as the final emotion class for the blog.

## 4. CONCLUSION

Emotion Detection can be seen as an important field of research in human-computer interaction. A sufficient amount of work has been done by researchers to detect emotion from facial and audio information whereas recognizing emotions from textual data is still a fresh and hot research area.

In this paper, methods which are currently being used to detect emotion from text are reviewed along with their limitations and new system architecture is proposed, which would perform efficiently.